\documentclass{elsart}

\usepackage{graphicx}
\graphicspath{{./img/}}
\usepackage{amsmath}
\usepackage{amssymb}
\usepackage{cite}

\renewcommand{\d}{\mathrm{d}}
\renewcommand{\i}{\mathrm{i}}
\renewcommand{\e}{\mathrm{e}}
\renewcommand{\L}{\mathrm{L}}
\newcommand{\R}{\mathrm{R}}
\renewcommand{\Re}{\mathop{\mathrm{Re}}}
\newcommand{\tr}{\mathop{\mathrm{tr}}}

\begin{document}

\begin{frontmatter}

\title{Coherent charge transport through molecular wires: influence of
strong Coulomb repulsion}
\author{Franz J. Kaiser},
\author{Michael Strass},
\author{Sigmund Kohler},
and \author{Peter H\"anggi}
\address{Institut f\"ur Physik, Universit\"at Augsburg,
Universit\"atsstra\ss e~1, D-86135 Augsburg, Germany}

\begin{abstract}

We derive a master equation for the electron transport through
molecular wires in the limit of strong Coulomb repulsion.  This
approach is applied to two typical situations: First, we study
transport through an open conduction channel for which we find that
the current exhibits an ohmic-like behaviour.  Second, we explore the
transport properties of a bridged molecular wire, where the current
decays exponentially as a function of the wire length.  For both situations, we discuss the differences
to the case of non-interacting electrons.

\end{abstract}

\end{frontmatter}

\section{Introduction}
\label{sec:introduction}

In recent years, it became possible to adsorb organic molecules via
thiol groups to a metallic gold surface and, thus, to establish a
stable contact between the molecule and the gold.  This opened the way
to reproducible measurements of the current through single molecules.
Experiments for such molecular conductance can be achieved in essentially two
ways:
One possible setup is an open break junction bridged by a molecule
\cite{Reed1997a, Kergueris1999a, Reichert2002a}.  There, the current
measurement provides evidence for \textit{single} molecule conductance because
asymmetries in the current-voltage characteristics reflect asymmetries
of the molecule \cite{Reichert2002a, Weber2002a}.
Alternatively, one can use a gold substrate as a contact and grow a
self-assembled monolayer of molecules on it.  The other contact is
provided by a gold cluster on top of a scanning tunnelling microscope
tip which contacts one or a few molecules on the substrate
\cite{Datta1997a, Cui2001a}.

Naturally, the experimental effort with such molecular wires is
accompanied by a vivid theoretical interest \cite{Nitzan2001a,
Hanggi2002elsevier, Nitzan2003a}.
Presently, the main theoretical focus lies on the \textit{ab-initio}
computation of the orbitals relevant for the motion of excess charges
through the molecular wire~\cite{DiVentra2000a,
Xue2002a, Damle2002a, Heurich2002a, Evers2004a}.

Another line of research employs rather generic models to gain a
qualitative understanding of the transport mechanisms involved.
The treatment of these models can be distinguished according to the
level at which interaction is taken into account.  Here, we are in
particular interested in two cases in which the many-body problem can
be traced back to the dynamics of single electrons on the wire:
The first case premises non-interacting electrons for which the
current can be computed from a Landauer-like formula
\cite{Mujica1994a, Segal2000a, FoaTorres2001a, Cizek2004a,
delValle2005a, Kohler2005a}.
The second case deals with the opposite limit in which Coulomb repulsion
is so strong that at most one excess electron can be
located on the molecule.  Such theories have been developed in the
context of conduction through coupled quantum dots \cite{Stoof1996a,
Brandes2004a, Flindt2004a} and for the incoherent transport through
molecular wires \cite{Petrov2001a, Petrov2002a, Lehmann2002a}.

In this work, we derive a master equation approach for molecular
conduction in the limit of strong Coulomb repulsion which restricts
the population of the molecular orbitals to zero or one excess electron.
Thereby, particular care will be taken in avoiding
inconsistencies like spurious non-vanishing transport in equilibrium
situations.
We present in Section~\ref{sec:model} our working model and derive in
Section~\ref{sec:current} a master equation which we evaluate for the two
mentioned limits, namely non-interacting electrons and strong Coulomb
repulsion.  Subsequently, we study in Sections \ref{sec:channel} and
\ref{sec:bridge} transport through open conduction channels and across
bridged molecular wires, respectively.  Explicit analytical
expressions for a wire that consists of only two sites are derived in
Appendix \ref{app:TLS}.

\section{Model}
\label{sec:model}

\begin{figure}[t]
  \centering
  \includegraphics[width=7.5cm]{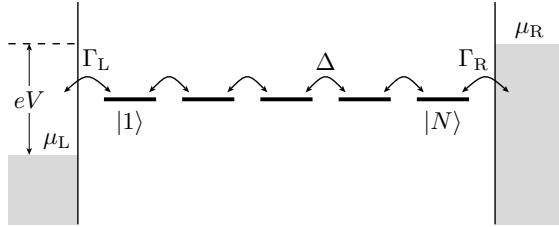}
  \caption{Tight-binding model for an open conduction channel with
  $N=5$ sites. An external bias voltage $V=(\mu_{\R}-\mu_{\L})/e$ is
  applied to the molecular wire.}
  \label{fig:model.channel}
\end{figure}
The setup at hand for studying coherent quantum transport is depicted
in Fig.~\ref{fig:model.channel}. The corresponding Hamiltonian reads
\begin{equation}
  \label{eq:H}
  H = H_{\mathrm{wire}} + H_{\mathrm{leads}} + H_{\mathrm{wire-lead}},
\end{equation}
where the individual terms describe the molecular wire, the electron
reservoirs of the leads and the coupling of the wire to the leads. The
wire itself is treated in a tight-binding approximation consisting of
$N$ orbitals. Since we aim at exploring blocking effects, the
corresponding wire Hamiltonian, incorporating the Coulomb repulsion in
the limit of a large interaction strength $U$, assumes the form
\begin{equation}
  \label{eq:Hwire}
  H_{\mathrm{wire}} =
      \sum_{n} E_{n} c_n^\dagger c_{n}
    - \Delta \sum_{n=1}^{N-1} \left(c_{n+1}^\dagger c_{n} +
      c_{n}^\dagger c_{n+1}\right)
    + U N(N-1).
\end{equation}
The fermion operators $c_n^\dagger$ ($c_{n}$) create (annihilate) an
electron in the orbital $|n\rangle$ and $E_{n}$ denotes the
respective on-site energy.  Here, we neglect the influence of the
voltage profile on the on-site energies \cite{Nitzan2002a, Pleutin2003a}. In the
Coulomb interaction term, $N=\sum_{n} c_n^\dagger c_{n}$ is the
operator counting the excess electrons
on the wire. The inter-site coupling characterised by the
hopping matrix element $\Delta$ is assumed to be equal between all
neighbouring sites. The leads attached to the molecular wire are modelled by ideal
Fermi gases,
\begin{equation}
  \label{eq:Hleads}
    H_{\mathrm{leads}} =
      \sum_q \sum_{\ell=\L,\R} \epsilon_{q} c_{\ell q}^{\dag}
      c_{\ell q} ,
\end{equation}
where $c_{\ell q}^\dagger$ ($c_{\ell q}$) creates (annihilates) an
electron with energy $\epsilon_{q}$ in lead $\ell=\L,\R$. As an initial
condition, we employ the grand-canonical ensemble of the electrons in
the leads at inverse temperature $\beta=1/k_{B}T$ and with
electro-chemical potentials $\mu_{\L/\R}$. Therefore, the lead
electrons are described by the equilibrium Fermi function
$f_{\ell}(\epsilon_{q})=\{1 + \exp[-\beta(\epsilon_{q} -
\mu_{\ell})]\}^{-1}$. For the initial density matrix, we then
have
\begin{equation}
  \label{eq:rholeadeq}
  \varrho_\mathrm{leads,eq}\propto
           \exp\left[{-\beta(H_\mathrm{leads}-\mu_\L N_\L -\mu_\R
           N_\R)}\right],
\end{equation}
where $N_{\ell}=\sum_{q} c^\dagger_{\ell q} c_{\ell q}$ denotes the
electron number in the left and right lead, respectively. From this
follows that all expectation values of the lead operators can be
traced back to the expression
\begin{equation}
  \label{eq:expectleadeq}
  \langle c_{\ell' q'}^\dagger c_{\ell q}\rangle =  \delta_{\ell\ell'}
  \delta_{qq'} f_{\ell}(\epsilon_{q}).
\end{equation}
The terminating sites $|1\rangle$ and $|N\rangle$, the so-called donor
and acceptor sites, couple via the
tunnelling matrix element $V_{\ell q}$ to the state $|\ell q\rangle$
in the respective lead. The Hamiltonian describing this interaction
has the form
\begin{equation}
  \label{eq:Hwire-lead}
  H_{\mathrm{wire-lead}} =
    \sum_{q} (V_{\L q} c^{\dag}_{\L q} c_{1} + V_{\R q} c^{\dag}_{\R
    q} c_{N}) + \mathrm{H.c.}
\end{equation}
It will turn out that the influence of the tunnelling matrix elements
is completely characterised by the spectral density
\begin{equation}
  \label{spectral.density}
  \Gamma_\ell(\epsilon) =
         2\pi\sum_q |V_{\ell q}|^2 \delta(\epsilon-\epsilon_q)
\end{equation}
which becomes a continuous function of $\epsilon$ if the lead modes are
dense.  If all relevant lead states are located in the centre of the
conduction band, the energy-dependence of the spectral densities is not
relevant and can be replaced by a constant, $\Gamma_\ell(\epsilon) =
\Gamma_\ell$.  This defines the so-called wide-band limit.

\section{Master equation approach}
\label{sec:current}

The computation of stationary currents can be achieved by deriving a
master equation for the dynamics of the wire electrons.  Thereby, the
central idea is to consider the contact
Hamiltonian \eqref{eq:Hwire-lead} as a perturbation.  From the
Liouville-von Neumann equation $\i\hbar \dot
\varrho=[H,\varrho]$ for the total density operator $\varrho$ one
obtains by standard techniques \cite{May2004a} the approximate
equation of motion
\begin{equation}
\label{mastereq}
\begin{split}
\dot\varrho(t)
= & -\frac{\i}{\hbar}[H_{\rm wire}(t)+H_\mathrm{leads},\varrho(t)] \\
  & -\frac{1}{\hbar^2}\int_0^\infty \d\tau [H_\mathrm{wire-lead},
     [\widetilde H_\mathrm{wire-lead}(-\tau),\varrho(t)]] .
\end{split}
\end{equation}
The tilde denotes operators in the interaction picture with respect to
the molecule and the lead Hamiltonian,
$\widetilde X(t)=U_0^\dagger(t)\,X\,U_0(t)$, where
$U_0$ is the propagator without the coupling.  For the evaluation of
Eq.~\eqref{mastereq} it is essential to use an exact expression for
the zeroth-order time evolution operator $U_0(t)$.  The use of any
approximation bears the danger of generating artifacts, which, for
instance, may lead to a violation of fundamental equilibrium
properties~\cite{Novotny2002a, May2004a}.

The stationary current defined as the net (incoming minus outgoing)
electrical current through contact $\ell$ is given by minus the
time-derivative of the electron number in that lead multiplied by
the electron charge $-e$, $I_\ell(t) = e(\d/\d t)\langle N_\ell\rangle$.
From the master equation \eqref{mastereq} follows
\begin{align}
I_\ell(t)
&= {}   e \tr[\dot \varrho(t) N_\ell] \nonumber
\\
&= {}
   -\frac{e}{\hbar^2} \int_0^\infty \d\tau \big\langle
   [\widetilde H_\mathrm{wire-lead}(-\tau),[H_\mathrm{wire-lead}, N_\ell]]
   \big\rangle ,
\end{align}
where we have used the relation $\tr A[B,C] = \tr[A,B]C$.
Next, we insert the wire--lead Hamiltonian \eqref{eq:Hwire-lead}, the
interaction-picture operator $\tilde c_{\ell q}(-\tau) = c_{\ell q}
\exp(\i\epsilon_q\tau)$ and the expectation values \eqref{eq:expectleadeq}.
By use of the spectral density \eqref{spectral.density}, the remaining
sum over the lead states is transformed into an integral which in the
wide-band limit $\Gamma_\ell(\epsilon) = \Gamma_\ell$ can be evaluated to
read
\begin{equation}
I_\ell
= \frac{e\Gamma_\ell}{\hbar} \langle c_1^\dagger c_1\rangle
   -e\frac{\Gamma_\ell}{\pi\hbar^2}\mathop{\rm Re}\int_0^\infty \!\d\tau \!
\int
   \d\epsilon\, \e^{\i(\epsilon+\mu_\ell)\tau/\hbar}
f(\epsilon)\big\langle[c_1,
   \tilde c_1^\dagger(-\tau)]_+\big\rangle .
\label{current.general}
\end{equation}

In the following, we specify the master equation \eqref{mastereq} and
the current formula \eqref{current.general} for
studying two limiting cases: The first limit $U=0$ describes non-interacting electrons.
For this situation, we follow the approach of
Ref.~\cite{Lehmann2002b}.
The second limit is the one of strong Coulomb repulsion in which
$U$ is much larger than any other energy scale of the problem.  Then,
only the states with at most one excess electron on the wire are relevant.

In both cases, a diagonal representation of the first term on the
right-hand side of the master equation \eqref{mastereq} is achieved by a
decomposition into the eigenbasis of the single-particle wire
Hamiltonian.  In this basis, the fermionic interaction picture
operators read
\begin{equation}
c_n(t) = \sum_\alpha \langle n|\phi_\alpha\rangle c_\alpha
\e^{-\i\epsilon_\alpha t} ,
\end{equation}
where $|\phi_\alpha\rangle$ denotes an eigenstate with energy
$\epsilon_\alpha$.  Below, we will need in particular the creation and
annihilation operators for the sites with direct contact to the leads,
i.e.\ $|n_\ell\rangle$ where $n_\L=1$ and $n_\R=N$.

\subsection{Non-interacting electrons}

In the limit $U=0$, the transport problem defined by the Hamiltonian
\eqref{eq:H} possesses an exact solution which is conveniently derived
within a scattering approach.  However, since one aim of the present
work is the comparison of two distinct master equations,
we only sketch the exact solution for the special case of a two-level system in
the Appendix \ref{app:tls.exact} and review here the corresponding
master equation approach \cite{Lehmann2002b, Kohler2005a}.

In general, the relation between the states $|\phi_\alpha\rangle$ and
the many-particle Hamiltonian \eqref{eq:H} is established via the
Slater determinant.  Alternatively, one can resort to Green's
functions.  In the present case, knowledge of the Green's function at
time $t=0$ is already sufficient. Apart from a prefactor, it is given
by the expectation value
\begin{equation}
P_{\alpha\beta}
=\langle c_\beta^\dagger c_\alpha\rangle =P_{\beta\alpha}^* .
\end{equation}
Then, one obtains from \eqref{current.general} for the stationary
current the expression
\begin{equation}
\label{current0}
I_0 = \frac{e\Gamma_\ell}{\hbar} \sum_\alpha\Big[
\sum_\beta
 \langle \phi_\beta | n_\ell\rangle\langle n_\ell|\phi_\alpha \rangle
 P_{\alpha\beta} -
  |\langle n_\ell|\phi_\alpha\rangle|^2
  f_\ell(\epsilon_\alpha)
\Big] ,
\end{equation}
where the index $0$ refers to $U=0$.
It can be shown that the current is independent of the index $\ell$, i.e.\
independent of the contact at which it is evaluated.  This reflects for a
two-probe setting the validity of the continuity equation.

In order to determine the expectation values $P_{\alpha\beta}$, we employ
the master equation \eqref{mastereq} and obtain for the stationary state
the condition
\begin{equation}
\label{master0}
\begin{split}
\i (\epsilon_\alpha -  \epsilon_\beta) P_{\alpha\beta}
=
\sum_{\ell=\L,\R} \frac{\Gamma_\ell}{2}
  \Big\{
  &
  \langle\phi_\alpha|n_\ell\rangle\langle n_\ell|\phi_\beta\rangle\,
  \big[ f_\ell(\epsilon_\alpha) + f_\ell(\epsilon_\beta) \big]
  \\
  - &
  \sum_{\alpha'}
  \langle\phi_\alpha| n_\ell\rangle\langle n_\ell|\phi_{\alpha'}\rangle\,
  P_{\alpha'\beta}
  -
  \sum_{\beta'}
  \langle\phi_{\beta'}| n_\ell\rangle\langle n_\ell|\phi_\beta\rangle\,
  P_{\alpha\beta'}
  \Big\} .
  \end{split}
\end{equation}
In a non-equilibrium situation, the solution of this set of equations
generally possesses non-vanishing off-diagonal elements, which in some
cases turn out to be crucial.

\subsection{Strong Coulomb repulsion}

In the limit of strong Coulomb repulsion, $U$ is assumed to be so
large that at most one excess electron resides on the wire.  Thus,
the available Hilbert space is restricted to the states
$\{ |0\rangle, c_\alpha^\dagger|0\rangle \}_{\alpha=1\ldots N}$, which we use for the
decomposition of the density operator to obtain
\begin{equation}
\rho = |0\rangle\rho_{00}\langle 0|
      + \sum_\alpha \big( c_\alpha^\dagger|0\rangle\rho_{\alpha0}\langle 0|
                         +|0\rangle\rho_{0\alpha}\langle 0|c_\alpha \big)
      + \sum_{\alpha\beta} c_\alpha^\dagger
      |0\rangle\rho_{\alpha\beta}\langle 0| c_\beta.
\end{equation}
With this ansatz, the current expectation value
\eqref{current.general} assumes the form
\begin{equation}
\label{current.inf}
I_\infty=e  \Gamma_\ell\sum_{\alpha} \big[ \sum_{\beta}
        \langle \phi_{\beta}| n_\ell \rangle \langle n_\ell | \phi_{\alpha}\rangle
        \bar{f}_\ell (\epsilon_{\alpha})
        \rho_{\alpha\beta}- |\langle\phi_{\alpha}| n_\ell \rangle |^2
        f_\ell(\epsilon_{\alpha})\rho_{00} \big] ,
\end{equation}
where $\bar f=1-f$.
The decomposition of the master equation \eqref{mastereq} into the
single-particle states $c_\alpha^\dagger|0\rangle$ provides for the stationary
state the set of equations
\begin{equation}
\label{masterinf}
\begin{aligned}
\i (\epsilon_\alpha - \epsilon_\beta) \rho_{\alpha\beta} =
\sum_{\ell=\L,\R}\frac{\Gamma_\ell}{2}\Big\{&
        \langle \phi_{\alpha}| n_\ell \rangle \langle n_\ell | \phi_{\beta}\rangle
                \big( f_\ell(\epsilon_\alpha)+f_\ell(\epsilon_\beta)\big)\rho_{00}
\\
-& \sum_{\alpha'}        \langle \phi_{\alpha}| n_\ell \rangle \langle n_\ell | \phi_{\alpha'}\rangle
        \bar{f}_\ell(\epsilon_{\alpha'})\rho_{\alpha'\beta}
\\
-&\sum_{\beta'}        \langle \phi_{\beta'}| n_\ell \rangle \langle n_\ell | \phi_{\beta}\rangle
        \bar{f}_\ell(\epsilon_{\beta'})\rho_{\alpha\beta'}
\Big\} .
\end{aligned}
\end{equation}
In order to fully determine the density operator, we need in addition
an expression for $\rho_{00}$ which can also be derived from the
master equation.  A more convenient alternative is given by the
normalisation condition $\tr\rho = \rho_{00}+\sum_\alpha
\rho_{\alpha\alpha} = 1$.  For the sake of completeness, we remark
that from the master equation~\eqref{mastereq} follows $\rho_{\alpha0} =
\rho_{0\alpha} = 0$ in the stationary state.

It can be shown that if the wire consists of just one site, i.e.\ for
$N=1$, both the master equation for $U=0$ and the one for $U=\infty$
provide identical expressions for the current.  The reason for this is
that already the Pauli principle inhibits the occupation of the
molecule by more than one electron.

\subsection{Rotating-wave approximation}
\label{sec:rwa}

For very weak wire--lead coupling, the coherent time-evolution
dominates the dynamics of the wire electrons.  This means that the
largest time-scale of the coherent evolution, given by the smallest
energy difference, and the dissipative time-scale, determined by
the coupling rates $\Gamma_\ell(\epsilon)$, are well-separated, i.e.,
\begin{equation}
  \label{rwa.condition}
  \Gamma_\ell \ll
  |\epsilon_{\alpha} - \epsilon_{\beta}|
\end{equation}
for all $\ell$ and $\alpha\ne\beta$.  Note that this condition is only
satisfiable if the energy spectrum has no degeneracies.  Then for
$\alpha\neq\beta$, the master equations \eqref{master0} and \eqref{masterinf},
which determine the stationary state, are dominated by their left-hand
side.  Consequently, $\rho_{\alpha\beta}$ is of the order
$\Gamma/(\epsilon_\alpha-\epsilon_\beta)$ such that it can be neglected
in the limit under consideration.  This constitutes the essence of a
rotating-wave approximation (RWA).  The above reasoning is equivalent to the
assumption that the stationary state is diagonal in the basis of the
eigenstates.  Within such a diagonal ansatz, it is possible to solve
both master equations analytically and, moreover, to provide a closed
expression for the respective stationary current.

\subsubsection{RWA for non-interacting electrons}

In the interaction-free case, the stationary state is found by
inserting the RWA ansatz $P_{\alpha\beta} = P_{\alpha\alpha}
\delta_{\alpha\beta}$ into equation \eqref{mastereq}; after some
algebra, we find
\begin{equation}
\label{RWA_solution}
P_{\alpha\alpha}
 = \frac{w_\alpha^\L f_\L(\epsilon_\alpha)
    +w_\alpha^\R f_\R(\epsilon_\alpha)}{w_\alpha^\L+w_\alpha^\R} .
\end{equation}
Thus, the populations are determined by an average over the Fermi
functions, where the weights
\begin{equation}
\label{weight}
w_\alpha^\ell = \Gamma_\ell|\langle n_\ell|\phi_\alpha\rangle|^2
\end{equation}
are given by the overlap of the eigenstate $|\phi_\alpha\rangle$ with
the site coupled to lead $\ell$.  Then the average current is readily
evaluated to read \cite{Lehmann2003b}
\begin{equation}
\label{RWAcurrent0}
I_{0,\mathrm{RWA}} = e\sum_{\alpha}
    \frac{w_\alpha^\L w_\alpha^\R}
     {w_\alpha^\L+ w_\alpha^\R}
    \big[ f_\R(\epsilon_\alpha) - f_\L(\epsilon_\alpha)\big] .
\end{equation}
This expression represents the limit $\Gamma\to 0$ of the
corresponding scattering theory \cite{Kohler2005a}.

\subsubsection{RWA for strong Coulomb repulsion}

The corresponding RWA ansatz for the strongly interacting limit reads
$\rho_{\alpha\beta}=\rho_{\alpha\alpha}\delta_{\alpha\beta}$.  Inserting it
into Eq.~\eqref{masterinf}, we find the solution
\begin{equation}
\rho_{\alpha\alpha}= \frac{1}{\mathcal{N}}\
\frac{w^\L_\alpha f_\L(\epsilon_{\alpha})+w^\R_\alpha f_\R(\epsilon_{\alpha})}
{w^\L_\alpha \bar{f}_\L(\epsilon_{\alpha})+w^\R_\alpha
\bar{f}_\R(\epsilon_{\alpha})} ,
\end{equation}
with the weight factors $w^\ell_\alpha$ defined as above and the
normalisation constant
\begin{equation}
\label{norm}
\mathcal{N}=
1+\sum_{\alpha'}\frac{w^\L_{\alpha'}
f_\L(\epsilon_{\alpha'})+w^\R_{\alpha'} f_\R(\epsilon_{\alpha'})}
{w^\L_{\alpha'} \bar{f}_\L(\epsilon_{\alpha'})+w^\R_{\alpha'}
\bar{f}_\R(\epsilon_{\alpha'})} .
\end{equation}
The average current follows directly by inserting
$\rho_{\alpha\alpha}$ into \eqref{current.inf} and reads
\begin{equation}
I_{\infty, \mathrm{RWA}} = \frac{e}{\mathcal{N}}
\sum_{\alpha} \frac{w^\L_\alpha w^\R_\alpha }
{w^\L_\alpha \bar{f}_\L(\epsilon_{\alpha})
+w^\R_\alpha \bar{f}_\R(\epsilon_{\alpha})}
\big[f_\R(\epsilon_\alpha)-f_\L(\epsilon_\alpha)\big] .
\end{equation}
This current formula differs from the one obtained within RWA for
non-interacting electrons, c.f.~Eq.~\eqref{RWAcurrent0}, by the appearance
of the normalisation factor $\mathcal{N}$ and by the Fermi functions
$\bar f = 1-f$ in the denominator.

\section{Open transport channel}
\label{sec:channel}

As a first application, we consider a wire for which
all on-site energies are at the level of the
chemical potentials and, moreover, all hopping matrix elements are
equal, as sketched in Fig.~\ref{fig:model.channel}.  Then, the molecular orbitals
are delocalised which provides ideal transport along the
molecule.  For a voltage which is sufficiently large such that all
molecular orbitals lie within the voltage window, the current is in
the interaction-free case dominated by the total transmission.
Under the assumption that all overlaps \eqref{weight} between molecule
eigenstates and sites $|n\rangle$ are identical, we find
$w_\alpha^\ell = 1/N$.  Therefore, the RWA current formula
\eqref{RWAcurrent0} becomes
\begin{equation}
\label{channel:I0}
I_0 = \frac{e\Gamma}{2\hbar} .
\end{equation}
In particular, we find that the current is independent of the wire
length which is characteristic for coherent transport of
non-interacting electrons through an open conduction channel~\cite{Datta1995a}.

This behaviour is significantly modified by the influence of strong
Coulomb repulsion.  The normalisation factor
\eqref{norm} reads $\mathcal{N}=N+1$  such that finally
\begin{equation}
I_\infty = \frac{e\Gamma}{\hbar (N+1)}.
\end{equation}
For $N=1$, this result coincides with Eq.~\eqref{channel:I0}, as
expected.  For a long wire, we find that the current decreases
$\propto 1/N$, i.e.\ with the inverse of the wire length.  This
behaviour resembles an ohmic resistor and has been observed in the
limit of strong Coulomb repulsion \cite{Lehmann2002a} also for
incoherent hopping of the wire electron.

\begin{figure}[tb]
  \centering
  \includegraphics[width=7.5cm]{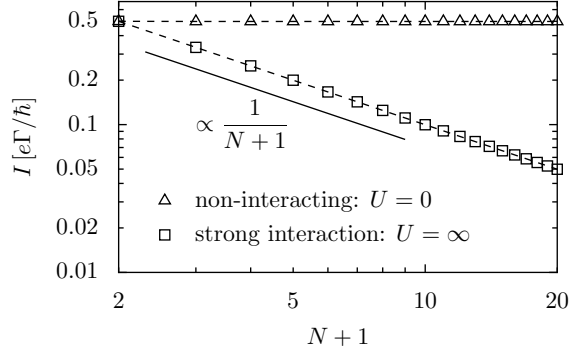}
  \caption{Stationary current as a function of the wire length for the
  transport through the open channel sketched in Fig.~\ref{fig:model.channel}
  with bias voltage $V=10\Delta/\hbar$.
  The other parameters are $\Gamma=0.1\Delta$ and $k_\mathrm{B}T=0.005\Delta$.
  The dashed lines are a guide to the eye.
  \label{fig:I.channel}}
\end{figure}%
The numerically computed current beyond RWA is shown in
Fig.~\ref{fig:I.channel}.  It fully confirms the respective length
dependence and, moreover, demonstrates the applicability of
the rotating-wave approximation in the present case.

\section{Bridged molecular wire}
\label{sec:bridge}

\begin{figure}[tb]
  \centering
  \includegraphics[width=7.5cm]{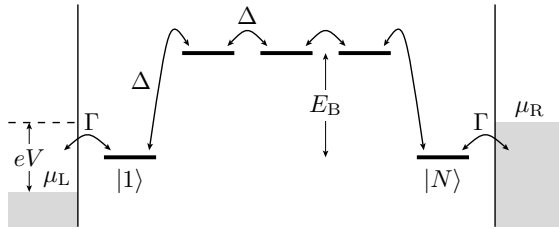}
  \caption{Level structure of the bridged molecular wire with $N=5$
  sites. The bridge levels are separated by $E_\mathrm{B}$ from the
  donor and acceptor levels $|1\rangle$ and $|N\rangle$.}
  \label{fig:model.bridge}
\end{figure}
Next, we consider the bridged molecular wire model sketched in
Fig.~\ref{fig:model.bridge}.  There, the energies of the donor and the
acceptor orbitals, $|1\rangle$ and $|N\rangle$, are assumed to be
close to the chemical potentials of the attached leads,
$\mu_\L \lesssim E_1 = E_N \lesssim \mu_\R$.  The bridge
levels $E_n$, $n=2,\dots,N-1$, lie $E_\mathrm{B}\gg\Delta$
above the chemical potential.

Let us first discuss the eigenstates of the molecule which discern
into two groups:
One group of states is located on the bridge.  It consists of $N-2$
levels with energies in the range $[E_\mathrm{B}-2\Delta,
E_\mathrm{B}+2\Delta]$.  In the absence of the driving field, these
bridge states mediate the super-exchange between the donor and the
acceptor.
The other group consists of the two remaining states.  They form a
doublet whose states are approximately
given by $(|1\rangle\pm|N\rangle)/\sqrt{2}$.  Its splitting can be
estimated in a perturbational approach and is
approximately given by $2\Delta(\Delta/E_\mathrm{B})^{N-2}$ \cite{Ratner1990a}.  Thus, the wire
can be reduced to a two-level system with the effective tunnel matrix
element $\Delta_\mathrm{DA}=\Delta\exp[-\kappa(N-2)]$, where
$\kappa=\ln(E_\mathrm{B}/\Delta)$.

The explicit calculation for the
two-level system is given in Appendix~\ref{app:TLS}.
Thus, in order to obtain the current for the present case, we just
have to replace in equations \eqref{TLS:I0} and \eqref{TLS:Iinf} the
tunnel matrix element $\Delta$ by $\Delta_{\mathrm{DA}}$.
For non-interacting electrons, we find from \eqref{TLS:I0} to lowest
order in $\Delta_\mathrm{DA}/\Gamma$ the expression
\begin{equation}
\label{bridge:I0}
I_0 = \frac{2e|\Delta|^2}{\hbar\Gamma}\mathrm{e}^{-2\kappa(N-2)} ,
\end{equation}
while in the case of strong Coulomb repulsion we employ
\eqref{TLS:Iinf} to obtain
\begin{equation}
\label{bridge:Iinf}
I_\infty = \frac{4e|\Delta|^2}{\hbar\Gamma}\mathrm{e}^{-2\kappa(N-2)} .
\end{equation}
In particular, one finds in both cases an exponentially decaying
length dependence of the current \cite{Mujica1994a, Nitzan2001a}.
Quite remarkably, the strong Coulomb repulsion enlarges the current by
a factor 2.

\begin{figure}[tb]
  \centering
  \includegraphics[width=7.5cm]{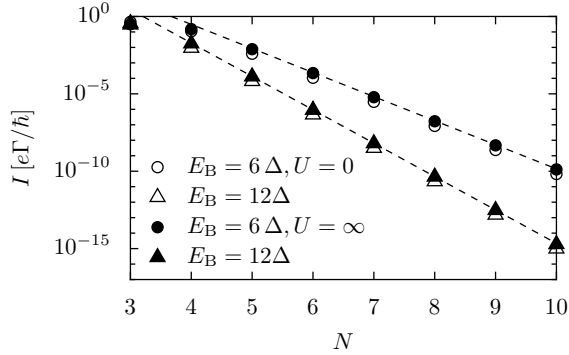}
  \caption{Current for the bridged molecular wire model, cf.
  Fig.~\ref{fig:model.bridge} comparing the non-interacting case with
  strong Coulomb repulsion.  The bias voltage is $V=5\Delta/\hbar$,
  $\Gamma=0.1\Delta$, and $k_\mathrm{B}T=0.005\Delta$.
  The dashed lines mark the analytical results for $U=\infty$.
  \label{fig:I.bridge}}
\end{figure}%
In order to test the quality of the two-level approximation above, we
compare the analytical result \eqref{bridge:Iinf}
against the numerical solution of the respective
master equation.  Figure \ref{fig:I.bridge} demonstrates the
almost perfect agreement between the numerical and the analytical
solution.  Moreover, it conforms the exponentially decaying length
dependence and the fact that the enhancement of the current by a
factor 2 owing to Coulomb interaction.  For $N=3$, the limit
$\Delta_\mathrm{DA}\ll \Gamma$ is not yet reached which explains the
small deviation from expression \eqref{bridge:Iinf}.

We close this section with the remark that for bridged molecular
wires, the rotating-wave approximation derived in Section
\ref{sec:rwa} results in $I_0=e\Gamma/2\hbar$ and
$I_\infty=e\Gamma/3\hbar$.  Thus, the RWA even fails to predict
qualitatively the
observed length dependence.  The reason for
this is that $\Delta_\mathrm{DA}\ll\Gamma$ and, thus, the condition
\eqref{rwa.condition} for the applicability of RWA is violated.

\section{Conclusions}

We have derived a master equation approach for the electron transport
through tight-binding systems in the presence of strong Coulomb
repulsion.  In contrast to prior work, we treat the master equation
beyond a rotating-wave approximation which extends the range of
validity of our approach to intermediately strong wire--lead coupling.
In particular, bridged molecular wires constitutes an example
for which our approach provides reliable results while within a
rotating-wave approximation one obtains qualitatively wrong results.

With this formalism, we have studied transport properties of two
models for molecular wires.  Thereby, we have worked out the
differences to the case of non-interacting electrons.
A model for which all on-site energies are identical, represents a
tight-binding version of an open conduction channel.  There, we
find a significant influence of Coulomb repulsion: While in the
absence of interaction, the current is length independent, it decreases
due to Coulomb repulsion proportional to the wire length.
Thus it resembles an ohmic conductor even though the transport is
fully coherent.

For the bridged molecular wire model, only the first and the last site
have energies close to the chemical potentials of the leads, whereas
all the other sites merely mediate co-tunnelling.  We
have demonstrated that then the wire exhibits the behaviour of a
two-level system with an effective tunnel matrix element.  In particular, we found the surprising effect that
Coulomb blocking enhances the current by a factor two.

Comparing the results for the open-channel model and the bridge model, we
can conclude that the influence of Coulomb repulsion depends
sensitively on the level structure of the molecule:  If
many unoccupied molecular orbitals have energies close to the chemical
potentials of the leads, electron-electron interaction reduces the
current considerably.

\section*{Acknowledgements}

The authors acknowledge financial by the Deutsche
Forschungsgemeinschaft through Graduiertenkolleg~283 and
Sonderforschungsbereich~486.
One of us (P.H.) would like to acknowledge many
elucidative, advisable and inspiring scientific discussions with
Philip Pechukas, who is still young and energetic enough to contribute
to great science and literature.

\appendix
\section{Two-level system}
\label{app:TLS}

The bridged molecular wire discussed in Section \ref{sec:bridge} can
be described by a conductor that consists of only a donor state
$|1\rangle$ and an acceptor state $|2\rangle$, i.e.\ $N=2$.  In this
appendix, we derive explicit results for the transport through an
unbiased two-level system ($E_1 = E_2 = 0$) for
$\Gamma_\L = \Gamma_\R = \Gamma$ and chemical potentials such that
effectively $f_\L=0$ and $f_\R=1$.  These two
sites are coupled by a tunnel matrix element $\Delta$.
Diagonalising the wire Hamiltonian~\eqref{eq:Hwire} for vanishing
Coulomb interaction ($U=0$), we obtain the bonding and anti-bonding
eigenstates and eigenenergies
\begin{equation}
\label{TLS:states}
\begin{split}
|\phi_+\rangle ={} & \frac{1}{\sqrt{2}} ( |1\rangle + |2\rangle),
\quad \epsilon_+ = -\Delta, \\
|\phi_-\rangle ={} & \frac{1}{\sqrt{2}} ( |1\rangle - |2\rangle),
\quad \epsilon_- = \Delta .
\end{split}
\end{equation}

For weak coupling between the donor and the acceptor, $\Delta \ll
\Gamma$, the master equation approach, albeit perturbational in
$\Gamma$, still provides the correct behaviour owing to the proper
inclusion of off-diagonal elements of the density matrix.  Within
rotating-wave approximation this is no longer the case.

\subsection{Landauer form}
\label{app:tls.exact}

According to Landauer~\cite{Landauer1957a}, the coherent transport for
non-interacting electrons can be interpreted as a quantum mechanical
scattering process. Thereby the in- and outgoing electronic states
scattered in the mesoscopic conductor are considered as plane waves.
As a consequence, the pivotal quantity which determines the system's
conductance is the transmission probability $T(E)$ and the
corresponding current can be written in the form
\begin{equation}
  \label{eq:Landauer}
  I_{0} = \frac{e}{2\pi\hbar} \int \d E  \,[f_{\R}(E) - f_{\L}(E)] T(E).
\end{equation}
The transmission can now be calculated via the relation
$T(E)=\Gamma_{\L}\Gamma_{\R} |G_{12}(E)|^{2}$, where $G(E) =
(E-H_\mathrm{wire} - \i\Gamma/2)^{-1}$ denotes the retarded Green's
function. For an unbiased two-level system, we obtain
\begin{equation}
  T(E) = \frac{\Gamma^{2}\Delta^{2}}{|(E-\i\Gamma/2) - \Delta^{2}|^{2}}.
\end{equation}
Inserting this expression into Eq.~\eqref{eq:Landauer}, one
arrives at
\begin{equation}
  \label{TLS:I0}
  I_0 = \frac{e\Gamma}{2\hbar} \frac{\Delta^2}{\Delta^2 + (\Gamma/2)^2} .
\end{equation}
A more explicit calculation can be found, e.g., in Ref.~\cite{Kohler2004a}.

\subsection{Non-interacting electrons}

For the eigenstates and eigenenergies \eqref{TLS:states}, the current
formula \eqref{current0}, valid for $U=0$, reads $I_0 = (e\Gamma/2\hbar)
\sum_{\alpha,\beta} P_{\alpha\beta}$ while the set of equations
\eqref{master0} becomes
\begin{align}
0 ={} & \frac{\Gamma}{2} ( 1 - 2P_{++} ) , \\
0 ={} & \frac{\Gamma}{2} ( 1 - 2P_{--} ) , \\
-2\i\Delta P_{+-} ={} & \frac{\Gamma}{2} ( 1 - 2P_{+-} ) .
\end{align}
This corresponds to $P_{++} = P_{--} = 1/2$ and $P_{+-} = \Gamma/(2\Gamma -
4\i\Delta) = P_{-+}^*$.  Inserting this solution into equation \eqref{current0},
we obtain for the stationary current the result \eqref{TLS:I0}.

The quality of the present master equation approach is underlined by
the fact that it here indeed reproduces even for $\Gamma \gg \Delta$ the
exact solution.  We emphasise that this is not the case for the RWA
solution \eqref{RWAcurrent0}: Since for this approximation by
definition $P_{+-}=0$, one obtains the result $I_{0,\mathrm{RWA}}
= e\Gamma/2\hbar$ which is independent of the inter-site
coupling $\Delta$.

\subsection{Strong Coulomb repulsion}

Using the eigenstates and eigenenergies \eqref{TLS:states}, one finds
that the current with Coulomb blocking, Eq.~\eqref{current.inf}, reads
$ I_\infty = (e \Gamma/2\hbar)\sum_{\alpha,\beta}\rho_{\alpha\beta}$.
Formally, this is identical to the corresponding expression for the
non-interacting case but with the $P_{\alpha\beta}$ replaced by the
matrix elements of the density operator in the basis of the single
particle states $c_\alpha^+|0\rangle$.  The stationary value of
these matrix elements is determined by Eq.~\eqref{master0} which for
a two-level system reads
\begin{align}
0 = {} & \frac{\Gamma}{2} (1 - 2 \rho_{++} -\rho_{--} - \Re\rho_{+-} ),
\\
0 = {} & \frac{\Gamma}{2} (1 -  2\rho_{--} -\rho_{++} - \Re\rho_{+-} ),
\\
-2 \i \Delta \rho_{+-} = {} & \frac{\Gamma}{2}
(1 -\frac{1}{2} \rho_{++} -\frac{1}{2}\rho_{--} + \rho_{+-}) ,
\end{align}
which corresponds to
\begin{align}
\rho_{++}=\rho_{--}={}&
\frac{8\Delta^2 +\Gamma^2}{2(12\Delta^2+\Gamma^2)},\\
\rho_{+-}=\rho_{-+}^*={}&-\frac{4\Delta \i \Gamma +\Gamma^2}{2(12\Delta^2+\Gamma^2)}.
\end{align}
Inserting into Eq.~\eqref{current.inf}, we finally obtain the current
\begin{equation}
\label{TLS:Iinf}
 I_\infty = \frac{e \Gamma}{2\hbar} \frac{2\Delta^2}{3\Delta^2 + (\Gamma/2)^2}.
\end{equation}

\end{document}